\documentclass[conference]{IEEEtran}
\IEEEoverridecommandlockouts
\usepackage{cite}
\usepackage{amsmath,amssymb,amsfonts}
\usepackage{algorithmic}
\usepackage{graphicx}
\usepackage{textcomp}
\usepackage{xcolor}
\usepackage{hyperref}
\usepackage{url}
\usepackage{multirow}
\usepackage{balance}

\newcommand{\gh}[1]{\emph{\textcolor{magenta}{G.H: #1}}}
\newcommand{\ma}[1]{\emph{\textcolor{blue}{M.A: #1}}}
\newcommand{\maisha}[1]{\emph{\textcolor{green}{maisha: #1}}}

\renewcommand{\gh}[1]{}
\renewcommand{\ma}[1]{}
\renewcommand{\maisha}[1]{}

\def\BibTeX{{\rm B\kern-.05em{\sc i\kern-.025em b}\kern-.08em
    T\kern-.1667em\lower.7ex\hbox{E}\kern-.125emX}}
\begin{document}

\title{A Survey on Congestion Control and Scheduling for Multipath TCP: Machine Learning vs Classical Approaches}
\author{\IEEEauthorblockN{Maisha Maliha}
\IEEEauthorblockA{\textit{School of Computer Science} \\
\textit{University of Oklahoma}\\
Norman, Oklahoma, USA \\
Email: maisha.maliha-1@ou.edu}
\and
\IEEEauthorblockN{Golnaz Habibi}
\IEEEauthorblockA{\textit{School of Computer Science} \\
\textit{University of Oklahoma}\\
Norman, Oklahoma, USA \\
Email: golnaz@ou.edu }
\and
\IEEEauthorblockN{Mohammed Atiquzzaman}
\IEEEauthorblockA{\textit{School of Computer Science} \\
\textit{University of Oklahoma}\\
Norman, Oklahoma, USA \\
Email: atiq@ou.edu}

}

\maketitle

\begin{abstract}
Multipath TCP (MPTCP) has been widely used as an efficient way for communication in many applications. Data centers, smartphones, and network operators use MPTCP to balance the traffic in a network efficiently. MPTCP is an extension of TCP (Transmission Control Protocol), which provides multiple paths, leading to higher throughput and low latency. Although MPTCP has shown better performance than TCP in many applications, it has its own challenges. The network can become congested due to heavy traffic in the multiple paths (subflows) if the subflow rates are not determined correctly. Moreover, communication latency can occur if the packets are not scheduled correctly between the subflows. This paper reviews techniques to solve the above-mentioned problems based on two main approaches; non data-driven (classical) and data-driven (Machine Learning) approaches. This paper compares these two approaches and highlights their strengths and weaknesses with a view to motivating future researchers in this exciting area of machine learning for communications. This paper also provides details on the simulation of MPTCP and its implementations in real environments.
\end{abstract}


\section{Introduction}
\ma{I have removed "Single path TCP" throughout the paper as TCP is always single path. i have also removed the SPTCP acronym which was not used anyway}

Communication is key in many domains, such as defense, hospitality, technology, and space. In telecommunications, packet switching is a method of grouping data in smaller packets for faster communication \cite{b0}. One of the earliest packet-switched networks started with the Advanced Research Projects Agency Network (ARPANET) \cite{b42} in the United States, which is also called the forerunner of the Internet. Today, the Internet has expanded and now consists of a set of protocols for global communications. Basically, a protocol is a set of rules that the sender and receiver must agree on to communicate with each other. 
The two well-known Transport layer protocols are User Datagram Protocol (UDP) \cite{b43} and Transmission Control Protocol (TCP). UDP does not need any handshaking, which means the receiver does not send any acknowledgment to the sender when it receives a message. UDP thus leads to faster communication.

TCP \cite{b40} provides more consistent communication by considering handshaking between the sender and receiver. With millions of devices connected to the Internet, there is a demand for faster communication, but TCP fails to meet that need. It's because of TCP's congestion control algorithm, which decreases the throughput (message delivery rate) in response to the loss of packets in the network. Also, the handshaking of TCP increases the time necessary for a packet to travel, resulting in higher latency. Keeping in mind these problems, Multipath TCP (MPTCP) has been introduced by the Internet Engineering Task Force (IETF) \cite{b63} to use multiple paths effectively and efficiently between the sender and receiver. 

TCP connections can experience packet losses or connection drops, resulting in a poor user experience \cite{b53}.  MPTCP can use multiple TCP connections, known as subflows, in parallel to overcome TCP's limitations.  One of the main goals of MPTCP is to control congestion and maintain traffic flows. Another focus of MPTCP is scheduling the packets over different sub-flows to send packets with the smallest round-trip time (RTT) \cite{b54}, which is the time taken to send a data packet to the destination and receive an acknowledgment from the receiver. There are a set of traditional techniques such as Dynamic-Window Coupling (DWC) \cite{b17}, Opportunistic Linked Increases Algorithm (OLIA)\cite{b14}, Balanced linked adaptation (BALIA) \cite{b15}, and Adaptive and Efficient Packet Scheduler (AEPS) \cite{b20} that control congestion or simultaneously schedule packets over multiple paths in MPTCP. 

Since the standardization of MPTCP, a lot of classical approaches have been proposed to improve the performance of the network in terms of throughput and latency, but most of them perform poorly in highly dynamic networks. Recently, \emph{data-driven} approaches, which are mostly based on Deep Reinforcement Learning, perform much better in dynamic networks because of their ability to learn the network conditions. To the best of our knowledge, there have been only a few proposals on controlling congestion while scheduling packets using deep reinforcement learning-based approaches. Although some methods have been proposed to control congestion, those works have not focused on reducing RTT. Some researchers have proposed schedulers to reduce latency, but they did not consider achieving high throughput.

One of the biggest benefits of using MPTCP is its capacity to use all the available subflows and boost the network's throughput. However, to achieve high goodput, a scheduling strategy is important.  Scheduling in MPTCP distributes packets over different subflows based on the smallest RTT. Schedulers using classical approaches, like \cite{b11, b12, b13}, increased the throughput but failed to adapt to the dynamic nature of a real-world network. They were tested using simulation, which, of course, does not emulate the real-world scenario. Machine Learning and 
Reinforcement Learning models have improved the above drawbacks as they can learn from past experience and they are more robust to the dynamic nature of real-world networks; However, machine learning techniques are usually slower than classical approaches and need a huge dataset to train the models \cite{b5, b23, b24}.

\subsection{Contributions}
The \textit{objective} of this paper is to provide a brief overview of the existing work on MPTCP, including both classical and machine learning-based approaches. We discuss how previous researchers have addressed MPTCP challenges and summarized their solutions. Previous works have reviewed the existing congestion control and scheduling techniques for MPTCP \cite{b54, b55,b56,b57,b59}. Among those works, some review papers have focused on either congestion control of MPTCP or only the scheduling of MPTCP, while others have reviewed only the existing work on MPTCP establishment. There are also some works that have mentioned both congestion control and scheduling but have not focused much on the scheduling-based works of MPTCP. 
The \textit{contributions} of this paper are as follows:
\begin{itemize}
    \item Discuss the difference between the traditional TCP and MPTCP communication protocols.
    \item Discuss in detail both the congestion control and the packet scheduling problems separately. 
    \item Comparison between the performance of ML-based and classical algorithms in terms of controlling congestion and packet scheduling.
    \item Summarize basic concepts in MPTCP, including the establishment of MPTCP in real-world platforms and simulators.
    \item Highlights advantages and limitations of previous works that can help the readers investigate future improvements on MPTCP. 
\end{itemize}

The rest of our paper is organized as follows; Section \ref{sec:background} describes the terminologies in communication and deep reinforcement learning. Section \ref{sec:mptcp_tcp} compares TCP and MPTCP. Sections \ref{sec:cc} and \ref{sec:schedule} describe previous works on congestion control and scheduling of MPTCP, respectively. Section \ref{sec:cc-schedule} focuses on the performance of both congestion control and packet scheduling. MPTCP implementations in the kernel and NS-3 are described in Section \ref{sec:implement}. Lastly, in Section \ref{sec:conclusion}, we conclude our survey by discussing future works in MPTCP congestion control and scheduling.   

\section{Background \& Terminologies}\label{sec:background}
\subsection{Overview of TCP}\label{sec:tcp}

TCP is a connection-oriented communication standard that computer applications use to communicate  over a network. It is a packet transfer protocol in the Transport Layer \cite{b73} of the TCP model. TCP uses only one dedicated path for packet transfer. Though TCP guarantees data integrity of the packets, it has to face packet loss, delay and other problems which are discussed in Section \ref{sec:mptcp_tcp}. Also, network congestion is another major problem in TCP which is discussed briefly later. Section~\ref{sec:some_concepts_of_tcp} summarizes some concepts in TCP which would be in common with MPTCP and they are also used in congestion control.

\begin{figure}[!h]
\centerline{\includegraphics[scale=0.60]{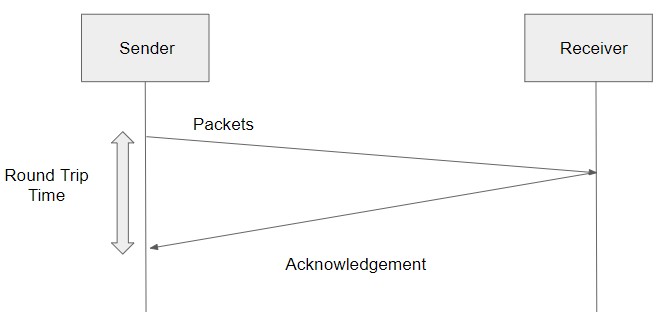}}
\caption{Illustration of the RTT of a packet from the sender to the receiver.} 
\label{fig:rtt}
\end{figure}

\begin{figure}[!h]
\centerline{\includegraphics[scale=0.50]{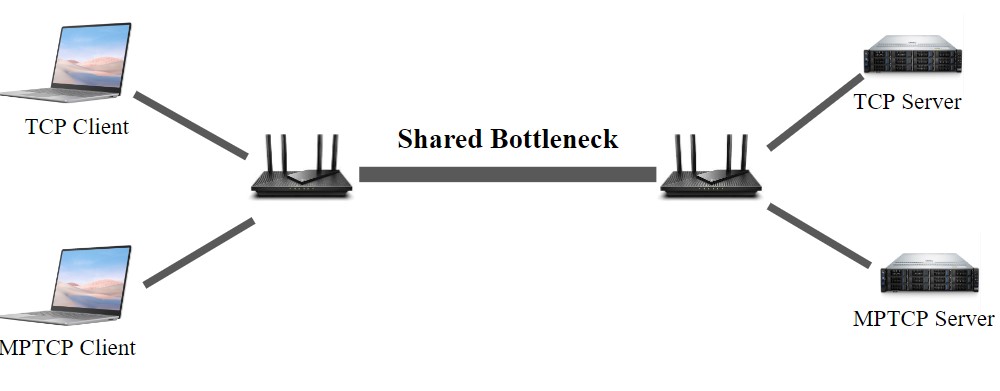}}
\caption{Illustration of shared bottleneck scenario in a network \cite{b25}.} 
\label{sha}
\end{figure}

\subsection{Some Concepts in TCP} \label{sec:some_concepts_of_tcp}
\begin{itemize}
\item \textbf{Round Trip Time (RTT):}
The time required to send a packet from the client to the server, and the time it takes for the server to receive an acknowledgment about receiving the packet is known as the round trip time (RTT). Reducing the round trip time is a primary focus of MPTCP. Figure~\ref{fig:rtt} illustrates the meaning of RTT.

\item \textbf{Throughput vs Goodput:} Throughput refers to the total number of packets transferred to the destination within a fixed time frame. On the other hand, Goodput is the number of meaningful packets that are delivered to the destination within a given time frame.

\item \textbf{Low Latency vs High Latency:} Latency refers to the amount of time required to send a packet from source to destination and back again. Low latency is always preferable.

\item \textbf{Congestion Window (CWND):}
The congestion window decides the number of bytes or how many packets will be sent at a given time. Depending on the larger congestion window size, the throughput will also be maximized. Congestion window size has been determined by the slow start and congestion avoidance phases of TCP which will be discussed in the next part.

\item \textbf{Bottleneck vs Shared Bottleneck:} Bottleneck occurs when there is not enough network capacity in a connection to handle the current volume of traffic. On the other hand, when a bottleneck link is shared between multiple subflows, it is referred as a shared bottleneck which is useful for maximizing the throughput. Figure~\ref{sha} illustrates the concept of a shared bottleneck scenario. 
 \gh{Atiq: could you please advise if the mentioned definition of bottleneck is clear?}

\end{itemize}

\subsection{Congestion Control in TCP} \label{sec:Congestion_control_in_TCP}
\begin{figure}[!h]
\centerline{\includegraphics[scale=0.42]{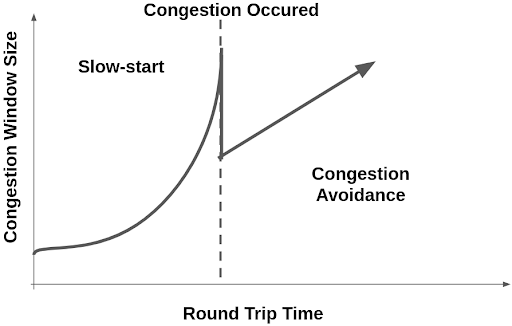}}
\caption{Change in the congestion window in AIMD algorithm when a packet loss is encountered.} 
\label{fig:aimd}
\end{figure}

TCP's congestion control mechanism has three phases; (1) slow start phase; (2) congestion avoidance phase; and (3) congestion detection phase. The basic difference between these three phases is the rate of increase in the congestion window size.
\begin{itemize}
\item \textbf{Slow Start Phase:}
Slow start phase works as a part of the congestion control algorithm in TCP by controlling the amount of data flow in a network. When a network becomes congested from excessive data in the network, the slow start phase chokes the traffic by limiting the congestion window size. In the slow start phase, the sender sends a packet that contains its initial congestion window, and the client responds with its maximum buffer size after receiving the packet. If the sender gets the acknowledgment from the receiver,  the number of packets to be sent to the receiver is doubled. This procedure continues until no acknowledgment is received. The acknowledgment may not be received for two reasons: if congestion occurs or the window limit of the client is reached. 

\item \textbf{Congestion Avoidance Phase- Additive Increase:}
 The congestion avoidance phase starts when the congestion window size of the TCP reaches a threshold in the slow start phase. In this phase, the size of the congestion window increases linearly. To elaborate, assume the congestion window size at time $t$ is 20 and all the packets have been transmitted successfully, then the congestion window size at time $t+1$ will be 21.
\item \textbf{Congestion Detection Phase- Multiplicative Decrease:}
If congestion occurs in the slow start phase or congestion avoidance phase, the congestion window size is decreased. This is called the multiplicative decrease phase, where TCP follows an exponential reduction of the congestion window. The additive increase and multiplicative decrease phases of the congestion avoidance and detection phases are referred to as Additive Increase Multiplicative Decrease (AIMD). An example of AIMD is shown in Figure~\ref{fig:aimd}.
\end{itemize}

\subsection{Overview of MPTCP}
\begin{figure}[!h]
\centerline{\includegraphics[scale=0.44]{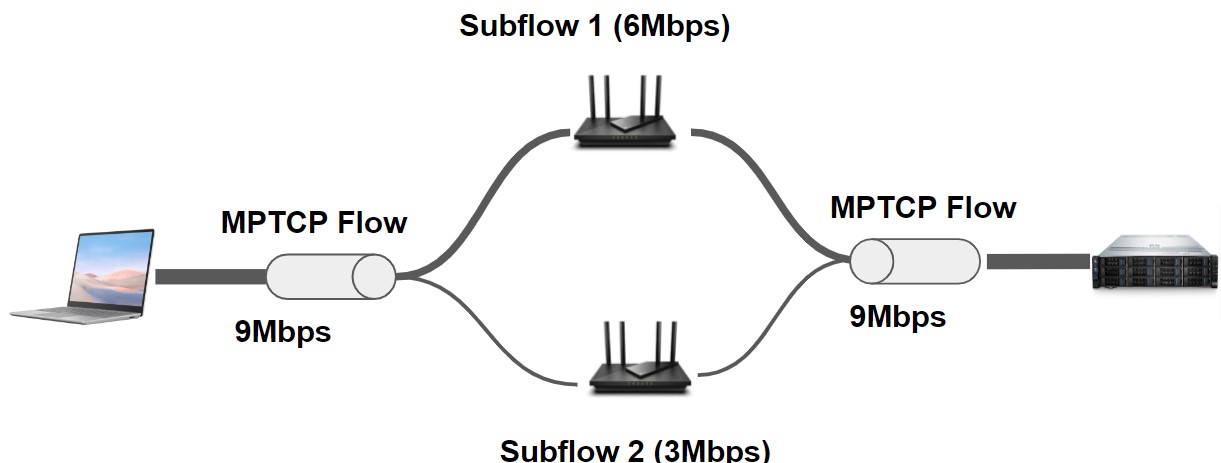}}
\caption{Overview of MPTCP}. 
\label{fig:mptcp}
\end{figure}

As opposed to TCP, which solely considers one path to transfer data, MPTCP is a transport layer protocol that allows the transfer of packets along multiple paths between the sender and the receiver. This helps the network increase its load capacity, thereby transferring a larger number of packets compared to TCP. The paths in MPTCP are called subflows. When one or more of the subflows fails to send a packet, it can flow through other subflows, leading to a fault-tolerant network. MPTCP is used in several areas of communication where there is a need for high throughput and very low latency during packet transfer. MPTCP is used in different applications such as online streaming, networking, gaming industries, VPNs. Figure~\ref{fig:mptcp} depicts the use of MPTCP where a cellphone may use either one of the subflows from two subflows to get connected to the server: one is the Wifi, and another subflow is the 5G network. The following section explains the procedure for establishing the MPTCP communication.
\subsubsection{Establishment of MPTCP Connection}
\begin{figure}[!h]
\centerline{\includegraphics[scale=0.5]{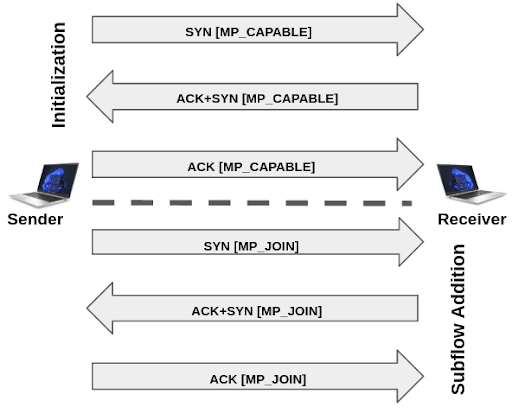} }
\caption{Establishment of an MPTCP connection.}

\label{fig:est}
\end{figure}

The establishment of MPTCP between a sender and the receiver has two stages; In the first step a single flow is established. This phase is similar to TCP. Then, subsequent subflows are created. In the first stage, the sender and the receiver use one subflow to set up the MPTCP connection between them by sharing randomly generated keys. This lays the foundation for creating further paths between the sender and the receiver. Figure ~\ref{fig:est} shows the establishment of an MPTCP connection using all subflows.

In MPTCP, after the establishment of the initial handshake as in TCP, the subsequent subflows are also handshaked \cite{b46}. MPTCP follows three-way handshaking consisting of SYN (synchronize), ACK+SYN and ACK (acknowledge). In the SYN packet, the sender shares its own token and a random nonce (number). Here the token is the hash value of the key using some cryptographic function which can be calculated in the initial phase with the keys exchanged in that phase. Subsequently, in the SYN+ACK packet, the receiver creates an HMAC (Hash-based message authentication code), the receiver's token and its nonce. 

The option MP\_CAPABLE is used to check whether the remote host is MPTCP enabled in the initial subflow. The option MP\_JOIN is used in the additional subflow establishment to associate with MPTCP connections. Lastly, the sender responds with its HMAC in the ACK packet. \cite{b46}

\subsection{Overview of Machine Learning Concepts used in Congestion Control in MPTCP}
\subsubsection{RNN and LSTM}
\begin{figure}[!h]
\centerline{\includegraphics[scale=0.33]{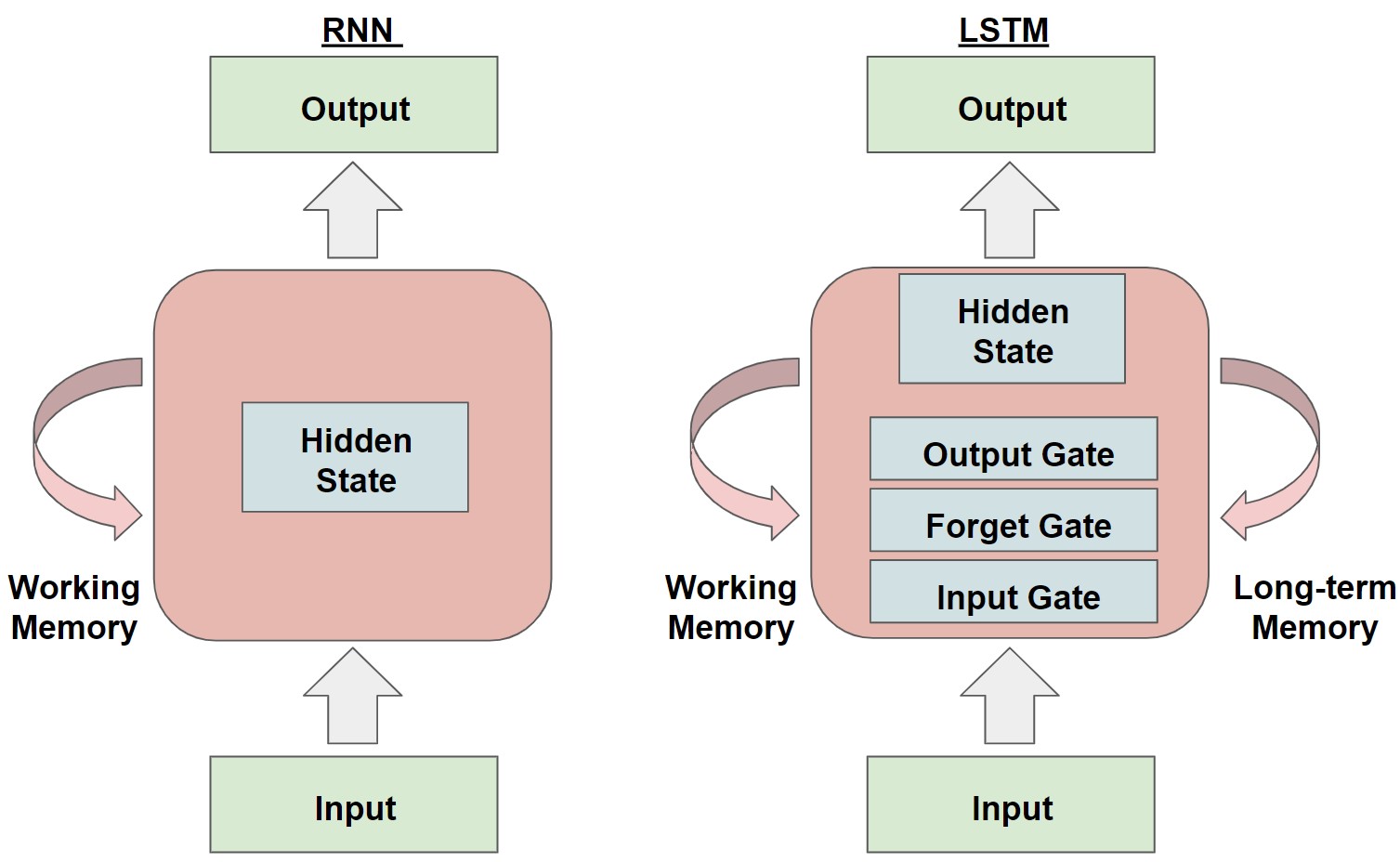}}
\caption{Architecture of RNN (left) vs LSTM (right).}
\label{lstm}
\end{figure}

RNN (Recurrent Neural Network) is one kind of neural network that passes the output from the previous steps to the next steps. RNN consists of the input layer, hidden layers, and the output layer. Its hidden state remembers the previous information to predict the next output. RNN works well in terms of correlation, while LSTM (Long Short-Term Memory) not only connects the correlation but also focuses on the context of the information. LSTM is an artificial neural network that follows feedback connections and stores previous information to predict the next one. It is a modified version of the RNN that solves the vanishing gradient problem \cite{b94} and can easily process longer sequences. LSTM has an input gate, an output gate and a forget gate. The Input Gate takes an input and vectorizes the input value. The forget Gate is responsible for forgetting unnecessary information, while the output gate generates the output. This helps the framework keep the necessary information and forget the unnecessary ones. Figure~\ref{lstm} shows the different components of an LSTM and compares with RNN. The LSTM-based framework is very popular to create Deep Reinforcement Learning-based congestion control system for MPTCP \cite{b2,b3}.

\subsubsection{Reinforcement Learning and Deep Q-Learning (DQN)}
A Markov model \cite{b68} consists of a tuple $(s,a,r(s,a))$, where $s$ is the current state vector, $a$ is the vector of actions an agent takes, and $R$ is the reward or the feedback the environment provides for the agent, given the current state and action. In a Markov Decision Process, the agent makes a set of actions, called policy, to maximize its expected reward. The function $Q(s,a)$ defines the optimal (\emph{i.e}., maximum) expected reward the agent can get, given the current state $s$ and action $a$. Evaluating $Q$ is important to plan for the optimal policy. Due to the stochastic nature of many environments and agents, calculating the exact value of $Q$ is usually not possible. Instead, the agent learns $Q$ values from its experience; this is called Q-learning, a type of reinforcement learning.

DQN Reinforcement Learning  (DRL) \ma{DRL was defined differently in a later section!! Use a different acronym here.}uses a deep neural network (DNN) to learn/estimate the $Q$ function. The deep neural network could be RNN, LSTM, CNN etc. In contrast, a traditional Q estimation (e.g., Temporal Difference (TD) learning \cite{b95}), is usually a memory table that stores all the previous records of the different steps that have been taken, along with their rewards which is not scalable for an environment with large size of space and actions, or when the action/state space is continuous. Deep-RL has been used in communication for learning the communication strategy between multiple agents. DRL is also used in MPTCP implementation to control congestion and schedule packets to maximize throughput and get low latency. In MPTCP communication network, the state can be throughput, sending rate, RTT, and loss of packets; actions should be increasing/decreasing congestion window size and reward is the measurement of either good or bad performance of the network, based on the state and action.



\subsubsection{Actor-critic model}
An actor-critic has two major components: an actor and a critic. An actor takes the state of the current environment and determines the best action that needs to be taken, depending on that state. Whereas the critic works for the evaluation role by taking the environment state with actions and returning a score/reward (\emph{e.g.,} Q) to decide how good is that action for that state. Determining the best action depends on the Q score, which is calculated separately in the critic network. An actor-critic learning technique learns both a policy function and a value function at the same time. The value function aids in enhancing the value function's training procedures, while the policy function instructs on how to make decisions.

\subsubsection{Transformers and Self-Attention}
The transformer is an encoder-decoder model and has been used in many applications \cite{b75}. Self attention follows the encoder part of transformers and can be used in many communication applications such as congestion control \cite{b3}, or scheduling packets for MPTCP. In natural language processing or machine translation, self-attention for a particular word measures the dependency of that word on the other words; more relevant words have a higher value in self-attention, and loosely connected words have lower values \cite{b27, b28}. This concept is brought into the area of communication and subflows. Here the words have been replaced by the subflows. In MPTCP, self-attention checks the dependency of a subflow with the other subflows by assigning different weights to the states of the other subflows. Here the states of the subflows may refer as RTT, packet loss, packet delay, and throughput \cite{b3}.

\section{Traditional TCP vs Multipath TCP}\label{sec:mptcp_tcp}
In this section, we discuss some major challenges in network communication, how MPTCP and TCP address them in different situations, including the comparison of MPTCP and TCP performance.
\subsection{Packet Loss}
Packet loss happens in a network when a packet fails to reach the receiver. Packet loss in a network indicates network congestion, disruption, or even a complete loss of connection. When a TCP connection encounters a packet loss, it considers a sign of network congestion and, therefore, halves the size of the congestion window and the threshold value; hence the throughput decreases. However, MPTCP is more robust in such cases; whenever it sees one of the subflows having packet loss, it reduces the congestion window of that subflow, while the other subflows remain intact. This was experimentally tested \cite{b34} where TCP and MPTCP transmission performances were compared using WiFi and LTE. In this real-world experiment, 750 individuals from 16 nations utilized a crowd-sourced smartphone application for 180 days. Deng et al. \cite{b34} compared MPTCP and TCP performance via WiFi or LTE from 20 different places across seven cities in the United States. The fastest link of normal TCP exceeds MPTCP performance for short flows. In this scenario, MPTCP fails to select an appropriate communication path to reduce transmission time for a small amount of data; hence, the packet loss increases for MPTCP. However, TCP experiences greater packet loss than MPTCP in log-distance communication experiments.
\subsection{Packet Delay}
Packet delay in a network is the time taken to transfer a packet of data from sender to receiver. The packet delay is highly affected by the number of routers or switches in the route, the nature of the path the packet follows, and the congestion in that path. The vulnerability of packet delay in both TCP and MPTCP scenarios can be described by one communication example. If a computer is connected via both wireless and wired connections, it may communicate with servers using TCP or MPTCP. In terms of TCP, if the connection is established over the wireless connection, it will have a greater overall delay than MPTCP. Wireless connection experiences high packet loss as opposed to the wired connection which leads to higher latency \cite{b60}. 

MPTCP has the choice to choose either a wired or wireless connection, it may transfer the packets via the wired connection if it experiences major packet delay through a wireless connection. Raiciu et al. \cite{b35} have compared the performance of MPTCP and TCP-based on the packet delays in different network topologies. Their findings reveal that MPTCP can achieve 90 percent bandwidth utilization and low overall packet delay when the number of subflows is two and eight in the VL2 \cite{b61} and FatTree network \cite{b71} topologies, respectively. The result also showed that MPTCP not only increases the bandwidth but also increases the robustness to network changes by lowering the packet delay.

\subsection{Out-of-Order Packets}
In TCP, a message is divided into multiple parts, known as packets. Each of these packets is given a unique number known as the sequence number. When these packets reach the receiver, it uses the sequence number to put them in order to retrieve the message. If these orders are not maintained during transmission, the delivery is called out-of-order delivery of packets. UDP is notorious for these deliveries as it does not consider any handshaking when each packet is received. But in TCP, the next packet is not sent until it gets the acknowledgment from the previous packet. If it gets a time out or negative acknowledgement (NACK) \cite{b62}, the packets are sent again; therefore, out-of-order delivery is rare in TCP. However, in MPTCP, there is a high chance of out-of-order delivery of packets as they use different subflows with different delays. Therefore, scheduling is one of the most challenging tasks in MPTCP, and a lot of work has been done to address this issue. Yang et al. \cite{b36} have mentioned a situation where the jitter can happen for transferring data in MPTCP. They tackled this issue using an innovative traditional scheduling process named Delay Aware Packet Scheduling technique to remove the jitter in the packets. Han et al. \cite{b24} used a queue to keep redundant packets that may get lost.

\subsection{Round Trip Time}
Round Trip Time (RTT) has been discussed in section \ref{sec:background}. 
Chen et al. \cite{b37} have compared the performances of TCP and MPTCP over WiFi and cellular networks where the authors compared the RTTs of the transmission protocols. They conducted two sets of experiments; in the first experiment, they used small-sized files, and in the second, large-sized files were used. In the first experiment, the file size varied from 8KB to 32MB. When WiFi is the default route, there is no discernible gain in MPTCP download performance over TCP. For small file downloads (such as 64 KB), the single route via WiFi delivers the optimum speed. However, a single LTE (Long-Term Evolution) \footnote{wireless broadband communication standard came before 4G network} channel becomes the optimum option for relatively longer traffic flows. MPTCP outperforms TCP for larger files.

 
\section{Congestion Control of MPTCP} \label{sec:cc}
Congestion control \ma{if you are using this acronym, this should be defined much earlier in the paper when congestion control is firt mention} is a concept of controlling congestion in a network and could happen in both TCP and MPTCP. Congestion occurs when there is too much data that needs to be sent through a network. Congestion control regulates the flow of data packets into the network, allowing for efficient use of a shared network infrastructure and preventing congestion collapse. In TCP, where there is only one subflow,  the network is easily congested.  

Different types of algorithms have been proposed to improve the congestion in TCP networks, such as TCP Cubic \cite{b66}, TCP Vegas \cite{b67}, and TCP Reno \cite{b67}. MPTCP provides several subflows, which results in a reduction of congestion. MPTCP has been designed to address the congestion issue, while still having the traffic flowing like a single-path TCP. A naive implementation of CC in a multipath setting would be using regular TCP congestion control for each subflow; however, it is not efficient as MPTCP which uses multiple concurrent TCP connections. Having a congestion control that manages the packet flows on subflows concurrently seems more efficient. For this purpose, many methods have been proposed to improve congestion in MPTCP. Congestion control algorithms for MPTCP are classified as classical and machine learning approaches which will be discussed in the following subsections.
\subsection{Classical Congestion Control Approaches}
Most of the existing congestion control algorithms in MPTCP setting focus on the Congestion-Avoidance (CA) phase that solely considers long flow transmissions and does not focus much on the slow start phase. The congestion avoidance phase prevents a network from being overflooded by data such that it discards packets with low priority to be delivered, and the rate of transmission rises linearly over time. Another approach is to focus on Slow Start. The Slow Start phase limits the quantity of data to be sent over a network to avoid congestion. However, it causes exponential growth of the congestion window in the uncoupled Slow-Start (SS) phase, leading to buffer overflow from burst data. In terms of solving the mentioned problems, Yang et al. \cite{b9} have proposed a Throughput Consistency Congestion Control (TCCC) algorithm which consists of both Coupled Slow-Start (CSS) and Aggressive Congestion Avoidance (ACA). The usage of CSS prevents packet loss brought on by large data bursts and ACA works on getting fair bandwidth which is shared in congestion avoidance. Their proposed framework enhances transmission efficiency. However, the CSS algorithm only plays a part in the initial slow start phase of MPTCP. As the subflows of MPTCP belong to different phases in congestion control (see Section \ref{sec:Congestion_control_in_TCP}), the CSS algorithm needs much extra consideration, which makes congestion control more challenging in MPTCP \cite{b9}. 

Traditional AIMD used in TCP shows poor performance adaptation in terms of network state-changing situations in MPTCP. Gilad et al. \cite{b10}, have presented a method named MPCC that uses online learning. Their implementation has been performed in Linux kernel and the method has been tested on different network conditions and many different network topologies. In terms of improving the implementation, their analysis needs to be reached beyond parallel link networks. As further research, they have mentioned about boosting the performance for short flows and solving the bandwidth mismatch problems on network paths. 

An energy-aware based congestion control algorithm (ecMTCP) has been developed by Le et al. \cite{b11} where the method distributes traffic between the most crowded and least crowded paths, as well as across paths with different energy costs, to achieve load balancing and energy savings. For simulation purposes, they used NS-2 simulator \cite{b76}, and their design mechanisms can work on getting higher throughput in terms of both TCP and MPTCP flows. The main goal was to shift the traffic to less energy-intensive and less crowded paths. Cao et al. \cite{b12} proposed weighted Vegas (wVegas), a delay-based congestion control scheme for MPTCP. This algorithm has detected the packet queuing delay of each path and ultimately has decreased the packet load of congested subflows by increasing the load of the less congested one. This framework performed traffic shifting which can cause less packet losses and provide better traffic balance in subflows. Cao et al. used NS-3 simulator \cite{b77} to conduct the simulation and build a Network Utility Maximization Model by proposing an approximate iterative algorithm to reach their aim of controlling congestion. 

Ji et. al \cite{b13} mentioned that existing multipath congestion control algorithms are unable to quickly adjust to dynamic traffic due to the heterogeneous Quality of Service (QoS). QoS refers the technologies that work on a network to manage traffic and enhance performance by reducing packet loss, delay, and latency in a network. It may lead to poor performance in certain network environments. To mitigate these issues, firstly, the authors have noticed the performance constraints of the most recent multipath congestion control algorithms through vast experimentation. Then, they used a unique control policy optimization phase referred to an adaptive QoS-aware multipath congestion management system that can quickly adapt to network changes. Their method uses the Random Forest Regressing (RFR) \cite{b70} method to carry out QoS-specific utility function optimization to adapt and encourage the improvement of the selected performance metric. They conducted the implementation in Linux kernel and showed their work outperformed most of the multipath congestion control methods such as wVegas\cite{b12}, Opportunistic Linked Increases Algorithm (OLIA)\cite{b14}, Balanced Linked Adaptation (BALIA)\cite{b15}. 

Singh et al. \cite{b16} improved the Opportunistic Linked Increases Algorithm \cite{b14} and Dynamic-Window Coupling (DWC) \cite{b17}. The authors provided a mechanism to reduce the overall packet reordering delay and focused on the buffer size in the receiver side. Their proposed work showed good performance in terms of various bottleneck scenarios. 

Hassayoun et al. \cite{b17} proposed a multipath congestion control scheme called Dynamic-Window Coupling (DWC) to obtain higher throughput to each end-to-end multiple paths. The authors also detected shared bottlenecks by monitoring loss and delay signals, and then grouped their congestion control mechanism over all subflows that shared a common bottleneck. Detecting the bottleneck for network conditions and regrouping the subflows in terms of the same bottlenecks leads to higher throughput. They also introduced subflow sets, a concept for enabling subflows to smoothly switch between independent and shared bottleneck-based congestion control. The algorithm has been implemented in NS-2 simulator. As future research, the authors introduced the possibility of including "memory" into the detection method for detecting the previous subflow groupings.

Ferlin et al.\cite{b18} have mentioned that increasing the bandwidth of multiple links and getting higher throughput will be impossible if two or more paths do not share a bottleneck. They found out the nonshared bottleneck paths of the coupled congestion control for links, they referred to it as a penalty, and to overcome it, they implemented shared bottleneck detection (SBD) algorithm for MPTCP. This work can balance congestion and throughput. Their observation has shown that in the case of non-shared bottleneck scenario, the maximum throughput can be achieved up to 40{\%} with two subflows. Also, the throughput gain increased by above 100{\%} when the number of subflows increased to five. Their implementation has been performed in Linux kernel and for emulation purposes, they have also used CORE network emulator \cite{b79}.

\subsection{Machine Learning Approaches for Congestion Control in MPTCP}
Though lots of work have been done in classical-based approaches for congestion control of MPTCP. But for controlling congestion, classical-based approaches focus solely on different types of congestion indicators (\emph{i.e.}, packet loss or RTT). In the case of classical-based approaches, the decision-making process totally depends on these unpredictable factors, which leads to poor performance. Whereas ML-based approaches aim to provide decisions based on experience, and can adapt well to any network situation. Thus, ML-based approaches outperform classical-based methods \cite{b84}.

In reality, networks are dynamic, and the state of the network changes frequently. Due to that, MPTCP performs poorly in many practical situations as MPTCP has to adapt in new network states. Zhuang et al.\cite{b1}, introduced a Reinforcement Learning technology that can learn the best route to send TCP packets such that the throughput has been maximized. They proposed a simple algorithm for controlling multipath congestion, where congestion control has been approached as a multi-armed bandit \cite{b65} issue based on online learning (MP-OL), which allowed flexible and adaptive transmission rate adjustments for each subflow with good performance. 

\begin{figure}[!h]
\centerline{\includegraphics[scale=0.48]{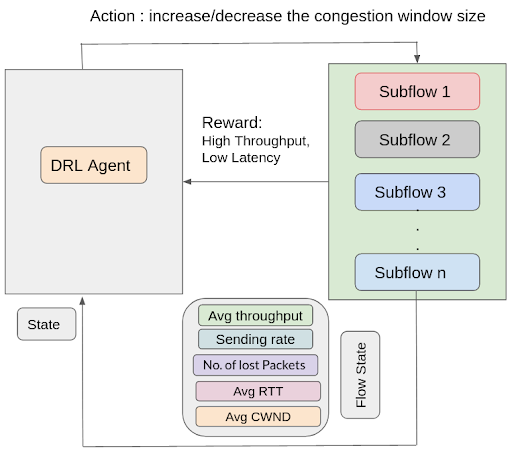}}
\caption{Subflows are controlled by a DRL agent.}
\label{fig:osdl}

\end{figure}

In \cite{b2}, the authors  proposed a Deep Reinforcement Learning (DRL)-based \ma{move this DRL definition to an earlier section as the term has been used in previous sections}framework to control congestion where a single DRL agent has been utilized to perform congestion control for all MPTCP flows to maximize the total utility. Figure~\ref{fig:osdl} illustrates the concept of DRL for MPTCP congestion control. They implemented the MPTCP in the Linux kernel and used an LSTM-based neural network under a DRL framework to develop a representation for all active flows. Their work was the first work where the authors incorporated the LSTM-based representation network into an actor-critic architecture for controlling congestion which used the deterministic policy gradient \cite{b78} to train the critic, actor, and LSTM networks. 

He et al. \cite{b3} worked on increasing/decreasing the sending rates of packets in response to congestion, where each DRL agent can control the congestion window size of each subflow. Their proposed DRL-based MPTCP framework also included self-attention, which has been used to check the dependencies of one subflow with the weighted sum of other subflows. They compared their work with DRL-CC \cite{b2} and showed their method outperforms DRL-CC. 

Li et al.\cite{b4} proposed a method called SmartCC which can learn a set of congestion rules for observing the environments and taking actions to adjust the congestion window size of each subflow. For the MPTCP implementation task, the authors used the NS-3 simulator. 

The Internet of Deep Space Things, or IoDST, offered communication services for mission spacecraft that send video data. To improve TCP throughput and stream playback, Ha et al. \cite{b5} designed a congestion control framework for MPTCP, which can be used for data streaming transmission. Their proposed Q-learning and Deep Q-Network (DQN)-based congestion control scheme calculated the ideal congestion window for data transfer in IoDST conversations. 

Xu et al. \cite{b6} proposed the SGIN-based High-Speed Railway (HSR) scenario with MPTCP. Space-ground integrated networks (SGINs) has been referred to as promising network architecture that provides seamless, high-rate, and reliable data transmission with incredibly wide coverage. By utilizing MPTCP in the SGIN, simultaneous data transfer over terrestrial and satellite networks has been made possible. However, due to MPTCP’s current congestion control (CC) mechanisms, it's difficult to know the difference between negative effects (like packet loss and/or increased round-trip time) brought on by congestion and those brought on by handovers. This may lead to severe performance degradation in the SGIN-based HSR scenario, where handover may occur frequently. To solve it, a DRL-based novel approach has been proposed to improve the goodput which outperformed other state-of-the-art algorithms. 
 
Xu et al.\cite{b7} presented a DRL-based novel framework for traffic engineering that can make decisions under the guidance of actor-critic networks. In their work, the state consisted of two components, such as the throughput and the delay of each communication session. On the other hand, the action has been defined as the solution to Traffic Engineering (TE) problems. The authors used the NS-3 simulator, and the reward of the model was the sum of the output from the utility function for an entire communication session. The utility function was a function of the throughput and delay of the network, which depicted how the network can perform. In the paper, each session had 20 iterations. In each iteration, the agent sent its actions to the environment and recorded the value from the utility function before updating the reward value. While they considered only one DRL agent (\emph{i.e.}, decision maker) in their framework, adding multiple agents can be considered to further improve the performance. 

Pokhrel et al. \cite{b8} introduced a transfer learning-based MPTCP framework for Industrial IoT, where the neighboring machines can collaborate to learn from each other. In their approach, when a new DRL system controlling the IoT network joins the environment, it can use the idea of transfer learning. \footnote{Transfer learning uses a previously trained model as the foundation for a new model on a different task.} NS-3 was used to simulate the algorithm. Their model has been proven theoretically and needs further research to determine its performance in a real-world situation.

 \section{Scheduling of MPTCP}\label{sec:schedule}

Scheduling of MPTCP decides the amount of data that needs to be scheduled to different subflows based on getting the higher performance (high throughput, low latency, less packet loss) in MPTCP. In this section, different classical and ML-based approaches have been discussed which can be used to schedule packets in MPTCP. 
 \subsection{Classical Approaches for Scheduling in MPTCP}
Hwang et al. \cite{b19} dealt with the problem of scheduling small-length packets. However, the authors mentioned MPTCP is usually advantageous for long-lived flows, and it performs worse than single-path TCP when the flow size is tiny (\emph{e.g.}, hundreds of KiloBytes). In this scenario, the quickest method is preferable since latency is far more critical than network bandwidth with such tiny data deliveries. The regular MPTCP packet scheduler may pick a slow path if the fast path's congestion window is unavailable, resulting in a delayed flow completion time.

To address this issue, Hwang et al. \cite{b19} suggested a novel MPTCP packet scheduler that momentarily blocks the slow path when the latency difference between the slow and fast paths is considerable, allowing the tiny quantity of data to be delivered swiftly via the fast path. The authors used the method to find the subflow with the lowest RTT regardless of the availability of the congestion window, and then they used the existing Lowest-RTT-First policy \cite{b83} to choose the optimal subflow. They then returned the best one if the difference between the best subflow RTT and the minimum RTT is less than a certain threshold. They picked 100ms for threshold delay when testing 3G and WiFi networks in this paper. 

Chaturvedi et al. \cite{b20}  analyzed different existing schedulers and identified some current outstanding concerns, such as head-of-line (HoL) blocking and out-of-order packet delivery. HoL blocking may occur when a single data packets queue may wait to be transmitted and the packet at the head of the line may not be able to move ahead due to congestion\cite{b82}. 
These problems reduce MPTCP performance and to mitigate the issues, the authors have presented an adaptive and efficient packet scheduler (AEPS). This novel MPTCP packet scheduler not only addresses these concerns but also offers high throughput with a short completion time by using the capacity of all available pathways. AEPS can send data packets to the receiver in the order they were received, and its performance is unaffected by the size of the receiver buffer, or the size of the data being transmitted. The AEPS has been developed with three objectives: (1) packets should arrive to the receiver buffer; (2) all pathways' bandwidth should be used; and (3) completion time should be as short as possible. According to the authors, the first condition assisted AEPS in resolving the HoL blocking and received window-limiting issues by sending packets to the receiver buffer in sequence. The second condition summed the bandwidth of each interface (path) by using all accessible pathways to the MPTCP source, which also helped to enhance throughput. The third criterion aided in choosing the routing for each packet so that the total network completion time can be minimized.

Dong et al. \cite{b21} thoroughly compared existing scheduling algorithms and guided the development of new scheduling algorithms in 5G. The authors examined the influence of several network parameters, such as RTT, buffer size, and file size, on the performance of current extensively used scheduling algorithms over a wide range of network circumstances. The paper compares the Lowest-RTT-First \cite{b83}, Delay-aware packet scheduler (DAPS) \cite{b91}, Out-of-order transmission for in-order arrival scheduler (OTIAS) \cite{b92} and Blocking estimation-based MPTCP scheduler (BLEST) scheduling \cite{b93} algorithms. The number of timeouts and flow completion time are compared in the path heterogeneity test. The results showed Lowest-RTT-First has the most timeouts, while the BLEST has the fewest. BLEST surpasses other algorithms in varying buffer size outcomes, followed by OTIAS and DAPS. Since BLEST can dynamically predict whether head-of-line blocking will occur and hence minimizes the quantity of out-of-order packets. In the different file size tests, BLEST and LowRTT perform better than DAPS, and OTIAS outperforms BLEST. 

Le et al. \cite{b22}  tackled the problem of out-of-order delivery in MPTCP. Because of the diverse nature of latency and bandwidth on each channel, the out-of-order packet issue becomes severe for MPTCP. To solve this issue, the authors presented the forward-delay-based packet scheduling (FDPS) method for MPTCP. The technique is divided into two parts: predicting the forward delay differences across pathways and picking data to send through a path when the congestion window is available.
\subsection{Machine Learning Approaches for scheduling in MPTCP}
In recent times, many ML-based approaches have been proposed to improve the scheduling mechanism of MPTCP. Though classical approaches achieve good performance in terms of scheduling in MPTCP, ML-based approaches also show promising result and becomes popular in terms of getting higher throughput with lower latency than non-ML methods.

Wu et al. \cite{b23} applied a learning-based technique to schedule packets in the different paths of an MPTCP. The authors have presented FALCON, a learning-based multipath scheduler that can adapt to changing network circumstances quickly and correctly using meta-learning. The meta-learning algorithm comprises two parts: offline training and online training parts. The online learning module captures the network-changing conditions whereas the offline learning module takes the experience(data) from the online module and divides the experience into different groups depending on the network conditions.

Han et al. \cite{b24} used the technique of redundancy of packets to reduce packet loss by suggesting EdAR (Experience-driven Adaptive Redundant packet scheduler). In the face of dramatic network environment changes, EdAR enables dynamically scheduling redundant packets using an experience-driven learning-based strategy for multipath performance enhancement. To allow accurate learning and prediction, a Deep Reinforcement Learning (DRL) agent-based framework has been created that learns both the network environment and the optimal course of action. EdAR has two transmission modes: standard transmission and redundant transmission. Standard transmission follows the regular data transition. Regarding the redundancy transmission, there is a buffer called redundant buffer. The redundant buffer holds packets that have already been transmitted but have yet to be acknowledged. If a new packet is transmitted from the send buffer on a subflow, it is copied to the redundant buffer. If a packet in the redundant buffer is not sent out or acknowledged, it is deleted from the redundant buffer. Silva et al. \cite{b86}, used linear regression \cite{b87} to predict throughput and latency in MPTCP subflows, and proposed Artificial Neural Network \cite{b88}-based linear classifier to choose the best subflow which can provide better performance in MPTCP scheduler. They implemented their work in NS-3 simulator.

\begin{table*}[htbp]
\centering
\caption{Performance measurement of reviewed papers for Congestion Control and Packet Scheduling in MPTCP. }

\begin{center}
\begin{tabular}{|c|c|c|c|c|c|}
\hline

\textbf{Approaches} & \textbf{Paper} & \textbf{\textit{Feature}} & \textbf{\textit{Strength}}& \textbf{\textit{Limitations}} & \textbf{\textit{Implementation}}\\ 

\hline

\multirow {24} {*} {ML} & DeepCC \cite{b3} & Congestion &increasing/decreasing the & increased computational  & Linux Kernel  \\
& & Control &  sending rates of packets& time & \\
& & &  &  & \\

\cline{2-6} 
& DRL-CC \cite{b2} & Congestion & high throughput & complexity by  & Linux Kernel \\
& & Control & &large state space & \\

\cline{2-6} 
 
& SmartCC \cite{b4} & Congestion & dealt with multiple communication & did not consider  & NS-3\\
&   & Control &  path in heterogeneous networks & TCP-friendliness   & \\
& &  &  & issues &  \\
 \cline{2-6} 
& IoDST \cite{b5} & Congestion &calculated the ideal congestion window & did not use real  & computer\\
& & Control &   & experiments or  &  simulations \\
& &  & & emulated tests & \\
\cline{2-6} 
& DRL for   & Congestion &higher goodput than other state-of-the-art &  & \\
&Handover-Aware & Control  &  algorithms & training time  & Linux Kernel \\
& MPTCP CC \cite{b6} & &  & becomes longer & \\
\cline{2-6} 
&MPTCP Meets   & Congestion & improved the efficiency of newly deployed& not proven in & NS-3 \\
&Transfer  & Control \& &   machines & practical situations &   \\
&  Learning \cite{b8} & Packet &  &   & \\
& & Scheduling & & & \\
\cline{2-6} 
&&&& &\\
&FALCON  \cite{b23} & Packet &can adapt in network changing conditions& needs to understand & Multipath  \\
& & Scheduling &   & the learning outcome & QUIC \cite{b85}\\

\cline{2-6} 
&EDAR  \cite{b24}  & Packet & enabled dynamically scheduling redundant & increased  & NS-3 \\
& & Scheduling & packets  &  computational & \\
& &  &  & time &  \\

\hline

\hline
\multirow {25} {*} {Classical}  & TCCC \cite{b9} & Congestion & improved efficiency in nonshared & needs extra   & NS-3  \\
&   & Control &  bottleneck scenario &  consideration for   & and Linux kernel\\
& &  &  & different phases  &   \\
&&  & & of MPTCP &   \\
\cline{2-6} 
&MPCC \cite{b10} & Congestion & tested on different network conditions& bandwidth mismatch   & Linux kernel \\
& & Control &  & problems on  & \\
&& & & network paths & \\


\cline{2-6} 
&ACCeSS  \cite{b13} & Congestion & quickly adjust to dynamic traffic& performance can be & Linux kernel\\
&& Control &   &  improved & \\


\cline{2-6} 
&CC MPTCP with  & Congestion   & balancing congestion with improving the & needs to improve & Linux kernel and \\
&shared bottleneck   & Control & throughput & it's robustness & CORE \\
& detection \cite{b18} &  &  && \\
\cline{2-6} 
&Packet scheduling   & Packet  & decreased the completion time of short & needs to improve & Linux kernel \\
&for multipath TCP \cite{b19} & Scheduling & flows  & overall transmission  & \\
&& &  &  rate &  \\
\cline{2-6} 
&AEPS  \cite{b20}  & Packet & high throughput and low completion & performed poorly &  Linux kernel \\
&  & Scheduling &  time  & in heterogeneous  & and NS-3 \\
&& & & networks & \\

\cline{2-6} 
&wVegas  \cite{b12} & Congestion & less packet losses & needs to effectively  & NS-3  \\

&& Control & & handle multiple  & \\
&& & & extended high-speed  & \\
&& & & paths &\\

\cline{2-6} 
&DWC for    & Congestion & high throughput & did not mention their  & NS-2 \\
&MPTCP-CC \cite{b17} & Control & & model performed   & \\
& & & & better than others & \\
\hline
\end{tabular}
\label{tab1}
\end{center}
\end{table*}

\section{Congestion Control and Scheduling of MPTCP} \label{sec:cc-schedule}
Few works have been done focusing on congestion control and scheduling the packets of MPTCP at the same time. Those works get higher throughput and lower latency in terms of performance evaluation. Though further research regarding congestion control with packet scheduling is needed to be done. This section reviews classical and machine learning approaches that have been done in this domain. 
\subsection{Classical Approaches on both Congestion Control and Scheduling of MPTCP} 
Wei et al. \cite{b25} proposed a model that gets higher throughput when the networks do not go through a shared bottleneck. Their work had two outcomes: (1) When no congestion would occur, their method has been able to get higher throughput than a single TCP. (2) When there is congestion in the network, their method has at least the same throughput as TCP. Their method also measured how severe or minor the congestion is in the network. They have introduced both SB-CC (Shared Bottleneck-based Congestion Control Scheme) and SB-FPS (Shared Bottleneck-based Forward Prediction packet Scheduling scheme), where SB-CC can detect shared bottlenecks and estimate the congestion degree of all subflows. SB-FPS can perfectly schedule data in shared bottleneck and can also distribute data according to the congestion window size of each subflows. For implementing MPTCP, they used the Linux kernel and achieved higher throughput. 

\subsection{Machine Learning Approaches on both Congestion Control and Scheduling of MPTCP}
Pokhrel et al. \cite{b26} have introduced the Deep Q learning (DQL)-based method to control congestion and schedule packets for MPTCP. Their proposed DQL framework has utilized the LSTM-based recurrent neural network where in their framework the Q function provided the logarithm value of goodput for the previous iteration. Here, the policy function was the actor-critic of two LSTMs and the value function was the reward. 
They considered RTT, throughput, and sending rate as the state. Depending on the state, their model provided action on whether window size needed to be increased or decreased and what changes can be taken in the schedule of packets for the subflows. In their work, the reward was the summation of all the Q functions for all subflows. Similar to other RL algorithms, the optimal decision was learned to maximize the reward. They have made their MPTCP implementation in the Linux kernel and achieved low delays with maximum goodput. 

\section{Implementation of MPTCP}\label{sec:implement}
In this section, we describe different ways of implementing MPTCP either in real hardware (kernel) or in simulator and list some of the works for each type of implementation as the reader's reference. Previously, NS-2 simulator has been used for implementing MPTCP \cite{b17,b96}. Now, most of the recent works have focused on implementing MPTCP in the Linux kernel after enabling MPTCP in the operating system or using the NS-3 simulator. Very few works implemented MPTCP on the CORE emulator. 

\subsection{Simulation}
Chihani et al. \cite{b29} implemented MPTCP in the NS-3 simulator and introduced a new protocol that worked better in various network conditions. They compared different packet reordering systems and analyzed that their implementations will be necessary for further MPTCP performance analysis in terms of controlling congestion. Nadeem et al. \cite{b30} worked on introducing three path managers; default, ndiffports, and fullmesh to create an MPTCP patch for implementing MPTCP in the NS-3 development version. While the default patch has not made any new subflows, fullmesh made a mesh of whole new subflows towards the feasible pairs of IP addresses; ndiffports introduced subflows in between the same IP pair with the help of distinct source and destination. It showed better results in terms of getting higher throughput and less flow completion time than prior works. Coudron et al. \cite{b31} proposed  MPTCP implementations in NS-3 to handle network traffic. They also compared their algorithm with previous work implemented in NS-3 and Kernel. Table~\ref{tab1} lists some other MPTCP implementations in NS-2, NS-3 and CORE simulator with the aim of congestion control or schedule packets or perform both congestion control and packet scheduling for MPTCP.

\subsection{Real Hardware (kernel)}
Network simulators sometimes fail to depict the original network conditions, as the real-world network is highly dynamic; the breaking of links and the creation of new links are spontaneous. Therefore, the evaluation of MPTCP on real-world networks using Linux kernels shows a much bigger picture of its strengths and weaknesses. In the work \cite{b44}, the authors have implemented MPTCP in a Linux kernel to study if each subflow has a different scheduler and then how the different subflows of an MPTCP may dispute bottleneck links with conventional single-path TCP. They tested LIA, OLIA, BALIA and wVegas on Linux kernel implementation of MPTCP and evaluated the throughput, latency, etc., on real-world networks.
Zannettou et al. \cite{b46} used the kernel implementation of MPTCP to show their MPTCP-aware scheduling performs better than random hashing of packets to subflows which is generally used. They used the FatTree \cite{b47} and Jellyfish \cite{b48} topologies to conduct their experiments. FatTree is a highly structured topology used in data centers to obtain the highest throughput cost-effectively, while Jellyfish is the most commonly used randomly structured topology which can support more hosts than the FatTree, while keeping almost the same throughput. The commercial application for MPTCP support is available online \cite{b80}. 

\section{Conclusion}\label{sec:conclusion}
This paper focuses on two crucial concepts in MPTCP - congestion control and scheduling. The study shows how the most recent works fulfill the previous work gaps and mitigate the above two MPTCP issues using different classical and ML-based approaches. Our study also presents the advantages and limitations of current works and encourages the researchers to continue further improvements in this domain. As in every communication sector MPTCP establishes a tremendous role, it is necessary to improve the performance of MPTCP, and our paper can be beneficial for the readers to have an extensive knowledge of MPTCP performance issues and can use it for proposing new algorithms.

\bibliographystyle{ieeetr}
\bibliography{ref}
\balance
\end{document}